\RequirePackage{fix-cm}
\documentclass{svjour3}                     % onecolumn (standard format)

\smartqed  % flush right qed marks, e.g. at end of proof
\usepackage{graphicx}
\usepackage{placeins}
\usepackage{siunitx}
\usepackage{amsmath}
\usepackage{url}

\urldef{\mailsa}[allowmove]\path|n.holzwarth, melanie.schellenberg, l.maier-hein]@dkfz-heidelberg.de|    

\begin{document}

\title{Tattoo tomography}
\subtitle{Freehand 3D photoacoustic image reconstruction with an optical pattern}

\titlerunning{Tattoo tomography}

\author{Niklas Holzwarth \and Melanie Schellenberg \and Janek Gröhl  \and Kris Dreher \and Jan-Hinrich Nölke \and Alexander Seitel \and Minu D Tizabi \and Beat P Müller-Stich \and Lena Maier-Hein \mailsa }

\authorrunning{Holzwarth, N. \textit{et. al.}}

\institute{Niklas Holzwarth \and Melanie Schellenberg \and Janek Gröhl \and Alexander Seitel \and Kris Dreher \and ~~~~~ Jan-Hinrich Nölke \and Minu D Tizabi \and Lena Maier-Hein 
\at Division of Computer Assisted Medical Interventions, German Cancer Research Center (DKFZ), Heidelberg, DE
\and
Lena Maier-Hein
\at Faculty of Mathematics and Computer Science, Heidelberg University, DE \newline Medical Faculty, Heidelberg University, DE
\and Beat Müller-Stich
\at Visceral and Transplantation Surgery, Heidelberg University Hospital, DE
}


\maketitle              % typeset the header of the contribution
\begin{abstract}\textbf{Purpose}:
Photoacoustic tomography (PAT) is a novel imaging technique that can spatially resolve both morphological and functional tissue properties, such as the vessel topology and tissue oxygenation. While this capacity makes PAT a promising modality for the diagnosis, treatment and follow-up of various diseases, a current drawback is the limited field-of-view (FoV) provided by the conventionally applied 2D probes. 

\textbf{Methods}: In this paper, we present a novel approach to 3D reconstruction of PAT data (\textit{Tattoo tomography}) that does not require an external tracking system and can smoothly be integrated into clinical workflows. It is based on an optical pattern placed on the region of interest prior to image acquisition. This pattern is designed in a way that a tomographic image of it enables the recovery of the probe pose relative to the coordinate system of the pattern. This allows the transformation of a sequence of acquired PA images into one common global coordinate system and thus the consistent 3D reconstruction of PAT imaging data.  

\textbf{Results}: An initial feasibility study conducted with experimental phantom data and \textit{in vivo} forearm data indicates that the \textit{Tattoo} approach is well-suited for 3D reconstruction of PAT data with high accuracy and precision. 

\textbf {Conclusion}: In contrast to previous approaches to 3D ultrasound (US) or PAT reconstruction, the \textit{Tattoo} approach neither requires complex external hardware nor training data acquired for a specific application. It could thus become a valuable tool for clinical freehand PAT.

\end{abstract}

\section{Introduction}

Photoacoustic imaging (PAI) is an emerging imaging modality that has undergone rapid development since the late 90s \cite{PA-review-wang}. PAI is based on the photoacoustic (PA) effect \cite{PAeffect} and - using a \textit{light in - sound out} principle - provides high contrast in optically absorbing tissue regions. Tissue is illuminated with nanosecond-long pulsed near-infrared laser light. When the light is absorbed by specific molecules, thermoelastic expansion causes a local pressure rise that generates an acoustic shock wave which can be measured by a standard ultrasound (US) transducer. Furthermore, acquiring multispectral PA data allows not only to spatially resolve highly concentrated absorbing tissue regions, but also to define the absorbing molecules and estimate their concentration in a process referred to as spectral unmixing \cite{SU-collagen}. It thus enables the measurement of both morphological and functional tissue information in depths up to several centimeters. A key application of PAI is the estimation of the blood oxygenation, which is relevant for a variety of diseases \cite{clinicalPAT}. \newline

Clinical PAI devices typically use handheld 2D ultrasound transducers for data acquisition. The bottleneck of such systems is that they produce only cross-sectional 2D slices of the imaged structures, although the full 3D context is crucial for various applications \cite{needle}\cite{epileptic}. Current hardware solutions towards 3D imaging include whole body scanners - which only allow small animal imaging - or 3D probes, which lack in penetration depth, field-of-view and spatial resolution and are associated with additional costs \cite{2d-3d-probe}. Another approach towards 3D US/PAI is based on constrained movement of a 2D probe, either introduced by additional mechanical hardware \cite{laurence} or light-weight robots \cite{robots}, which also offers additional intrinsic tracking data. An alternative are software-based solutions for compounding 3D volumes based on pose estimation of the individual 2D image slices. However, these may also rely on complex hardware, such as an external tracking system \cite{Kirchner}. Mobile cameras mounted to the probe for motion tracking have also been investigated \cite{6dof} but come at the cost of a complex hardware setup as well. To overcome these issues, sensorless approaches based on speckle decorrelation \cite{chen_1997}\cite{gao_2016} have been proposed for US, but they are not well applicable to PA images, which do not show speckle effects due to the physical difference in image formation between PAI and US \cite{wang_speckle}. Recent work enhanced the speckle-based approach with deep learning based components \cite{prevost}. However, machine learning-based approaches rely on training algorithms for specific applications and may thus lack in robustness when transferred to new settings. \newline

To overcome the limitations of prior work, we present an entirely novel approach to 3D reconstruction that leverages the unique capability of PAI to tomographically reconstruct optical properties. Specifically, the concept is based on an optical pattern placed on the region of interest which allows the recovery of the 3D probe pose relative to the pattern coordinate system based on a single tomographic image. By transforming a sequence of freehand PAI images into the common pattern coordinate system, a consistent 3D reconstruction of PAI imaging data can be obtained without the need for an external tracking system. \newline 

The following sections present the first prototype implementation of the novel approach (patent pending) along with a feasibility study assessing the accuracy and precision of the reconstruction.

\section{Material and Methods} \label{sect:mat}

In this section, (1) the general concept of \textit{Tattoo tomography} is outlined, followed by a description of the first prototype implementation, comprising information on (2) the used PAI imaging device, (3) the first optical pattern designed for the approach, (4) the corresponding approach to pose estimation and (5) the method for compounding. Finally, (6) the validation concept is described.

\subsection{Tattoo tomography concept} % Tattoo concept +++++++++++

\begin{figure}[h!tb]
    \centering
    \includegraphics[trim = 0mm 0mm 0mm 0mm, clip, width=0.85\linewidth]{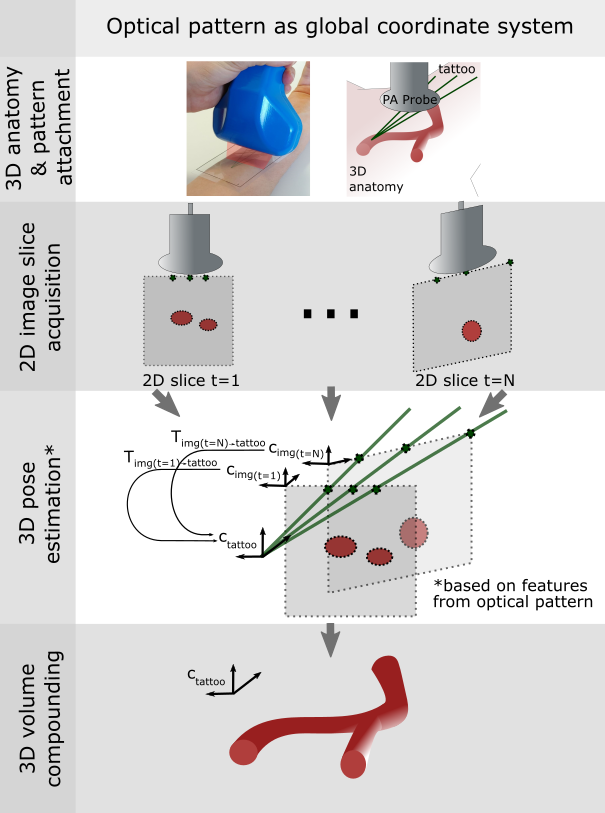}
    \caption{\textit{Tattoo tomography}: The approach to 3D image reconstruction comprises four steps. (1) Prior to image acquisition, the optical pattern is placed on the region of interest. (2) Next, a sequence of images is acquired, each showing a part of the pattern as well as the target region. (3) The image features corresponding to the pattern (here: three points) are extracted from each slice and used to recover the pose of the probe relative to the pattern coordinate system. (4) Once all acquired slices have been transformed (T) into a common coordinate system (C), the 3D volume is compounded.}
    \label{fig:tattoo-principle}
\end{figure} 

The process of reconstructing a 3D image with the \textit{Tattoo tomography} concept is illustrated in Fig. \ref{fig:tattoo-principle}. 
\begin{enumerate}
    \item Pattern attachment: Prior to image acquisition, an optical pattern is placed above the anatomical region of interest. This pattern should fulfill the following key requirements:  (1) It should be designed in a way that a tomographic image of it enables the estimation of the probe pose relative to the coordinate system of the pattern and that  (2) it is easy to mount and remove in a clinical setting. (3) The dye that makes up the pattern (here seen in green) should absorb in the frequencies that match those of the imaging system (here: near-infrared) and that are complementary to those used for the actual imaging.  
    \item Image acquisition: A sequence of images is acquired, each showing a part of the pattern as well as the target region. 
    \item Pose estimation: As the pose information is encoded in the optical pattern by design, the pose of each image slice within the pattern coordinate system can be determined via the PA image features corresponding to the pattern.
    \item Image compounding: The 3D volume is reconstructed by interpolating between the transformed slices in the pattern coordinate system.
\end{enumerate}

The following sections present the first prototypical implementation of this concept.

\subsection{PA imaging device} All PA images used as part of this work were acquired with a multispectral optoacoustic tomography (MSOT) Acuity Echo research system (iThera Medial GmbH, Munich, Germany). It is a hybrid device, which can acquire co-registered and synchronised PA and US images. The built-in laser is a Nd:YAG laser with a tuning range from 660 -- 1300 nm, a peak pulse energy of 30 mJ, a repetition rate of 25 Hz and a pulse duration of 4 -- 10 ns. The concave 2D US transducer has a 4 MHz center frequency, 256 elements and a radius of 40 mm.

\subsection{Optical pattern} % Prototype implementation
For this first prototypical implementation, a trident design was chosen, as depicted in Fig.~\ref{fig:c-tattoo}. This trident consists of an isosceles triangle (for simplicity referred to as \textit{tilted lines}) and the corresponding angle bisector between the tilted lines of the triangle (referred to as \textit{central line}). 

Mathematically speaking, this initial prototype pattern design currently only allows accounting for three degrees of freedom (DoF), representing the intersection line of the image coordinate system with the \textit{Tattoo} coordinate system. The remaining three DoFs are recovered by imposing constraints on the image acquisition process: Firstly, the optical pattern has to be approximately placed such that it resembles a flat surface, and secondly, the PA probe is required to be positioned orthogonally to the pattern plane (\textit{orthogonality constraint}).

The pattern was printed on a transparent foil commonly used in overhead projection (PrintLine Overheadfolie, Vleveka, Deurne, Netherlands) using cyan ink of a print station (TASKalfa 5052ci, Kyocera, Esslingen, Germany). These foils are well-suited for skin contact, easily sterilizable and printable in a straightforward manner. During acquisition, the pattern provides a strong contrast at a wavelength of \SI{750}{nm} while the functional imaging was performed at a wavelength of \SI{850}{nm}, where oxygenated hemoglobin is a dominant absorber in tissue. The foil is placed on tissue with a thin layer of coupling gel on top and underneath the foil and fixated using tape to avoid movement of the foil relative to the skin. The setup allows to easily place and remove the pattern on / from the tissue of interest. \newline

\subsection{Pose estimation}  \label{sec:pose}
\begin{figure}
    \centering
    \includegraphics[trim = 0mm 0mm 0mm 0mm, clip, width=.9\linewidth]{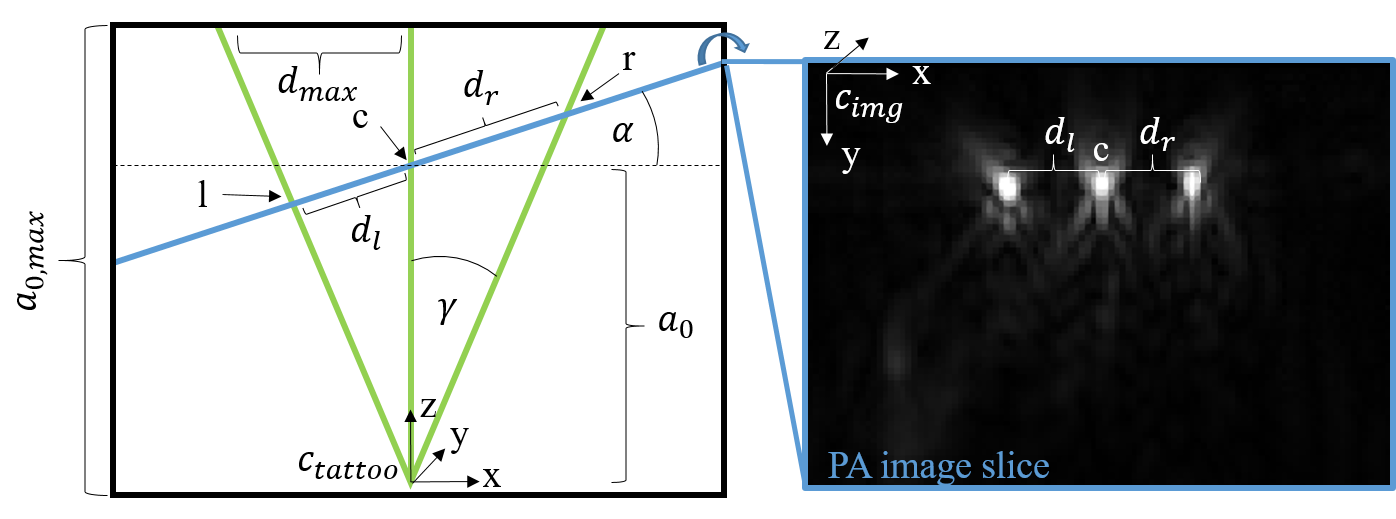}
    \caption{2D representation of the tattoo containing all relevant variables (left) and corresponding photoacoustic image (right). It shows the pattern (green), and the intersection of the image slice with the pattern (blue line). The image clearly shows three points representing the part of the image that intersects with the optical pattern (l, c, r). Based on the distances $d_l$ and $d_r$ between neighbouring points, the pose of the intersection line, represented by a blue line can be unambiguously determined. 
    \label{fig:c-tattoo}}
\end{figure}

The pose estimation step refers to estimating the pose of an individual image slice relative to the pattern coordinate system ($C_{tattoo}$). To this end, the information corresponding to the optical pattern is extracted from the PAI image that reflects the wavelength in which the pattern is absorbing (here: 750 nm). When using the optical pattern proposed in the previous section, this task is equivalent to extracting the three intersection points of the image with the line-based pattern, as depicted in Fig.~\ref{fig:tattoo-principle}. This is achieved with the SciPy \cite{Scipy} function \textit{find\_peaks}. Based on the three points, the line corresponding to the probe pose on the optical pattern (blue line in Fig.~\ref{fig:c-tattoo}) can be unambiguously determined. It is characterized by the distance $a_0$ of the absorption peak of the central pattern line \textit{c} to the pattern origin and the slice tilt angle $\alpha$ as defined in Fig.~\ref{fig:c-tattoo}. These are computed as follows:

\begin{equation}
    tan(\alpha) = \frac{1}{tan(\gamma)} \cdot \frac{d_l - d_r}{d_l + d_r},
    \label{eq:alpha_acc}
\end{equation}

\begin{equation}
    a_0 = \frac{(d_l+d_r)  \cdot cos(\alpha) + \frac{(d_l-d_r)^2}{d_l+d_r}}{2 \cdot tan(\gamma)}, 
    \label{eq:ao_acc}
\end{equation}

where $d_r$ and $d_l$ are the distances between the right and left point to the central point respectively and $\gamma$ is the pattern opening angle. When assuming an orthogonal pose of the probe relative to the pattern plane, the line given by $\alpha$ and $a_0$ uniquely determines the pose of the probe in 3D. This results in the following transformation matrix $T_{img\xrightarrow{}tattoo}$ which transforms the PA image slice into the tattoo coordinate system: 

\begin{equation}
    T_{img\xrightarrow{}tattoo}= \begin{pmatrix}
    cos(\alpha) & 0 & sin(\alpha)   & -x_c \\
    0           & 1 & 0             & -y_c \\
    -sin(\alpha)& 0 & cos(\alpha)   & a_0  \\
    0           & 0 & 0             & 1
    \end{pmatrix},
    \label{eq:transformation}
\end{equation}

where $x_c$ and $y_c$ are the x and y coordinates of the central \textit{Tattoo} point $c$ in the PA image slice and $\alpha$ and $a_0$ the previously derived \textit{Tattoo} coordinates. 

\subsection{3D volume compounding} \label{sec:compounding} 

The 3D compounding algorithm is inspired by the "Volume reconstruction algorithm" of the public software library for ultrasound imaging research (PLUS) \cite{plus}. A 3D zero-valued voxel grid is defined inside the bounding box of all transformed slices. The voxel spacing is set corresponding to the pixel spacing of the image in x-y direction and predefined according to the probe movement in z direction. Each image pixel in each slice is assigned to the closest voxel location. The individual voxel values are then calculated as the average pixel value of all assigned pixels.

\subsection{Experimental setup} \label{sec:exp} %++++++++++++++++++++++++++++++++++++++
The purpose of our experiments was to assess the general feasibility of the \textit{Tattoo tomography} concept. We provide (1) a qualitative visual demonstration, (2) a quantitative validation of the accuracy in a phantom experiment and (3) a quantitative validation of the precision in an \textit{in vivo} experiment.

The pattern was constructed with an opening distance $d_{max}$ of 20 mm and a vertical extension ($a_{0,max}$) of 50 mm. The scans of all experiments were performed using the two wavelengths 750 nm and 850 nm, where 750 nm corresponds to the wavelength in which the optical pattern is absorbing strongly (and thus well-visible) and 850 nm corresponds to wavelength used for the construction of the "functional" tissue information. The PA data was reconstructed using a delay-and-sum \cite{DAS} algorithm. 

\paragraph{Feasibility demonstration:} For demonstration of the principal feasibility of the approach, a logo of the international conference on Information Processing in Computer Assisted Interventions (IPCAI) was printed and placed inside a gel pad (AQUAFLEX Ultrasound Gel Pad, Parker Laboratories, inc., Fairfield, USA), that was cut in half in height (see Fig.~\ref{fig:IPCAI} a \& b). The optical pattern was fixed on top of the gel pad with pins. For scanning, the probe was moved along the pattern in two principal ways: (1) Slow and without direction changes, representing a careful acquisition and (2) fast with many direction changes, representing an extremely careless acquisition (see Fig.~\ref{fig:bad}, left column).

\begin{figure}
    \centering
    \includegraphics[trim = 0mm 0mm 0mm 0mm, clip, width=.8\linewidth]{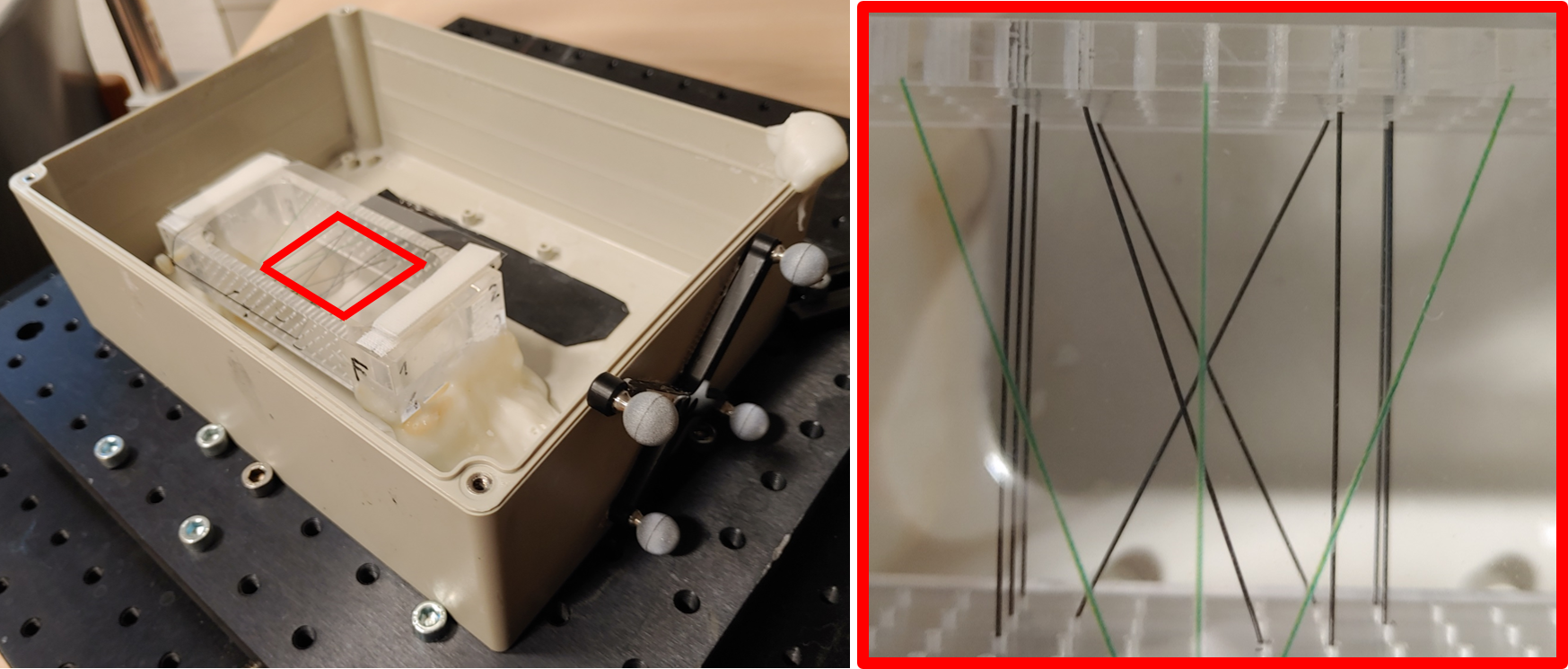}
    \caption{Phantom used for the quantitative validation of reconstruction accuracy. The optical markers (silver spheres) enable the comparison to a baseline 3D reconstruction method. The red box (right: zoomed in)  highlights the optical pattern placed on top of the N-wires. The wires have a diameter of \SI{0.4}{mm} and the holes of the frame are \SI{5}{mm} apart from each other.}
   \label{fig:phantom_exp}
\end{figure}

\FloatBarrier
\paragraph{Phantom validation:}%+++++++++++++++++
The phantom used for the phantom validation was an N-wire phantom (see Fig.~\ref{fig:phantom_exp}) typically used for hand-eye calibration of tracked devices \cite{Kirchner}\cite{plus}. The phantom was built using three layers of N-shaped black nylon wires (0.4 mm in diameter) showing a strong broadband absorption in the near-infrared. It was placed within a milk bath containing 1.5 l of tap water and 20 g of coffee cream (10 \% fat) to increase the scattering coefficient at room temperature. The phantom wiring was transferred to a 3D model based on the known location of the wires. A point representation of this model was then used as reference for the validation, as detailed below (see Fig. \ref{fig:ICP}).

For image acquisition, the optical pattern was fixed on top of the phantom using Velcro strips. This setup was placed upon a movable optical breadboard, and the probe was clamped above the phantom with a fixed probe angle $\alpha$ (rotation around y-axis). To enable the comparison with a baseline 3D reconstruction method, the probe was tracked with an optical tracking system (NDI Polaris Spectra camera, NDI Medical, Waterloo, Canada). During the scanning, the breadboard was moved under the probe. The PAI scans were then compounded using both the \textit{Tattoo} method (cf. section~\ref{sec:compounding}) as well as the optical tracking baseline method described in \cite{Kirchner}. Subsequently, a 3D segmentation of the PA wire volume was performed by applying a background suppressing threshold using the Medical Image Interaction Toolkit \cite{mitk}. 

Based on the known N-wire model and the 3D reconstructions, the accuracy of the compounded images was assessed. As the relation of the phantom coordinate system and the \textit{Tattoo} coordinate system can only be approximated, we concentrated the validation on the consistency of the reconstructed model. Specifically, we applied the iterative closest point (ICP) algorithm as implemented in the MITK \cite{A-ICP} framework (see Fig.~\ref{fig:ICP}) to register the reconstructed point cloud with a (simulated) point cloud representing the N-wire model. The fiducial registration error (FRE) was then used for quantifying the accuracy of the 3D reconstruction (see Table~\ref{tab:ICP-phantom}). As stopping criterion for the ICP algorithm, the change of the FRE between two ICP iterations had to be smaller than 0.001 mm.

\FloatBarrier
\paragraph{\textit{In vivo} validation:}%+++++++++++++++++
In the \textit{in vivo} validation step, we concentrated on validating the precision (reproducibility) of our method. To this end, we obtained a total of 30 \textit{Tattoo} scans from three healthy volunteers. For each volunteer, the optical pattern was placed on the forearm, and 10 measurements were taken while tracking the PAI probe with an optical tracking system. All measurements were transformed to the \textit{Tattoo} coordinate system (for our method) and the coordinate system of the optical tracking system (for the baseline method), and two corresponding 3D reconstructions were performed. The largest image region, corresponding to a vessel and comprising measurements of at least 7 volumes, served as target region for the assessment. To approximate precision, we performed a slice-based analysis. The reference position of the vessel center was computed as the mean of the (up to) 10 vessel locations extracted from the individual scans. The latter were again determined with the \textit{find\_peaks} function (see Sec.~\ref{sec:pose}). The mean Euclidean distance between this reference and the individual measurement served as metric for the precision of the compounding and referred to as mean vessel distance (MVD).

\FloatBarrier

\section{Results} %++++++++++++++++++++++++++++++++++++++

\begin{figure}
    \centering
    \includegraphics[trim = 0mm 0mm 0mm 0mm, clip, width=\linewidth]{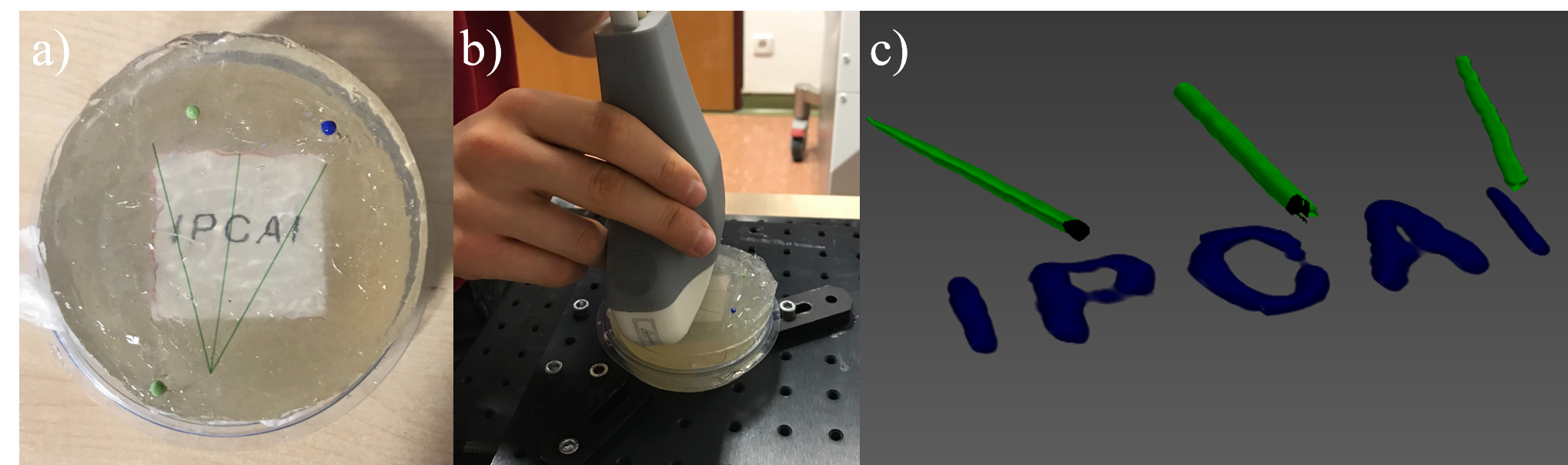}
    \caption{The experimental setup and results of the \textit{Feasability demonstration} experiment. The IPCAI logo printed on paper is placed underneath a gel pad on top of which the optical pattern is fixed and some ultrasound gel is added (a). A freehand PA scan is acquired (b), which results in a fully image-based and clear reconstruction of the IPCAI logo.}
   \label{fig:IPCAI}
\end{figure}

\begin{figure}[h!]
    \centering
    \includegraphics[trim = 0mm 0mm 0mm 0mm, clip, width=\linewidth]{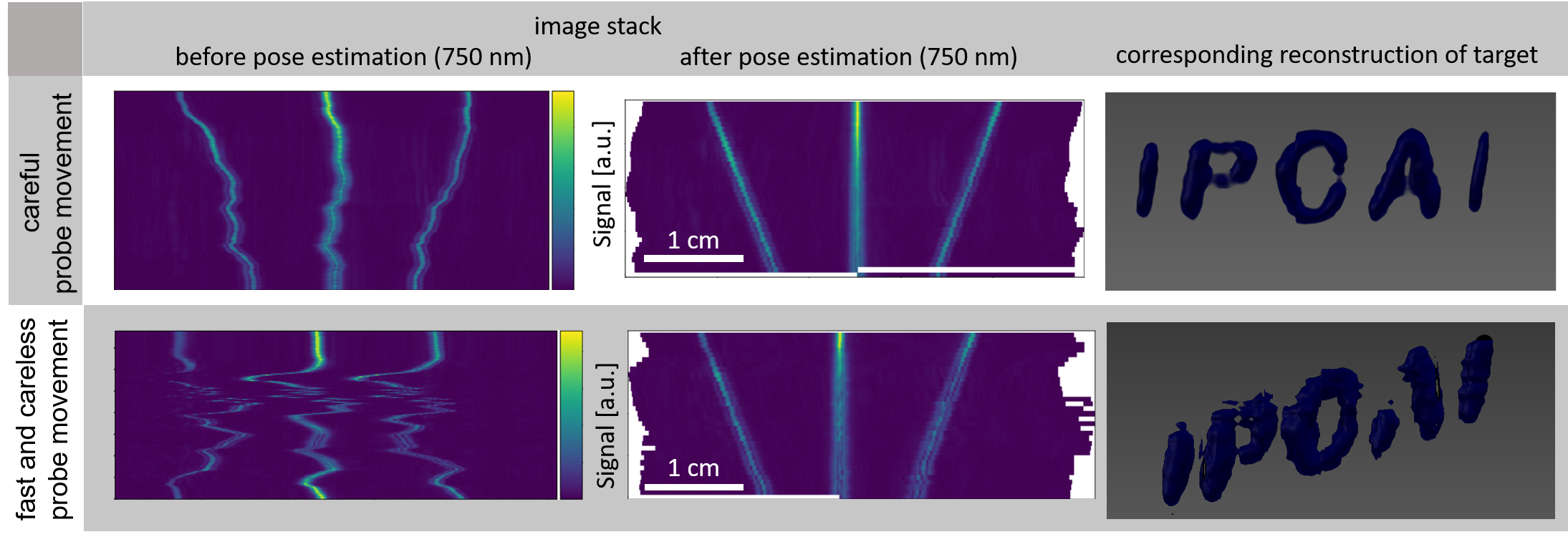}
    \caption{Good and poor reconstruction of a target region (here the text IPCAI, right) corresponding to a relatively slow and careful image acquisition (top) and to a fast and careless acquisition (bottom). The 2D images represent an intersection of the image stack at a fixed y-position before (left) and of the 3D volume after (middle) \textit{Tattoo} reconstruction.}
    \label{fig:bad}
\end{figure}

\paragraph{Feasibility demonstration} %+++++++++++++++++
\FloatBarrier
Fig.~\ref{fig:IPCAI} and Fig.~\ref{fig:bad} show the setup and the results for the feasibility experiment. It can be seen that the \textit{Tattoo} reconstruction is feasible even when images are acquired in an extremely careless and unsystematic manner. 

\begin{figure}[h!]
    \centering
    \includegraphics[trim = 0mm 0mm 0mm 0mm, clip, width=\linewidth]{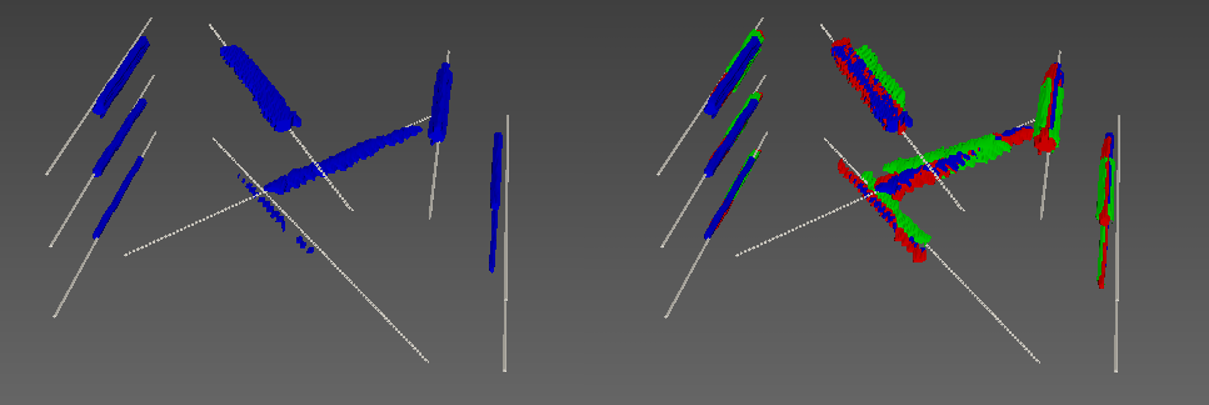}
    \caption{The 3D model of the N-wire phantom (white) is displayed together with the optical pattern compounded volumes of a single scan (blue, left) and all three scans (blue, red, green, right) after applying an ICP algorithm to register the \textit{Tattoo} point cloud on the N-wire model (cf. Table~\ref{tab:ICP-phantom}).}
    \label{fig:ICP}
\end{figure}

\FloatBarrier
\paragraph{Phantom validation:}%+++++++++++++++++
\FloatBarrier
The results of the \textit{Tattoo}-compounded volumes registered with a 3D model of the N-wire phantom can be seen in Fig.~\ref{fig:ICP}. The FREs for the registration of the compounded volumes (\textit{Tattoo} and optical tracking) with the N-wire model are shown in Table \ref{tab:ICP-phantom} for three different, but fixed, rotation angles $\alpha$ (0°, 4°, 8.5°). It is 0.63 mm on average for the \textit{Tattoo} compounding and 0.87 mm on average for the optical tracking-based compounding. 

\begin{table}[h!]
    \centering
    \caption{Consistency of 3D reconstruction of the proposed approach (\textit{Tattoo}) and the baseline method (optical tracking), assessed by registering a model of the phantom wires with the corresponding reconstructions (see. Sec.~\ref{sec:exp}). The fiducial registration error (FRE) of the iterative closest point (ICP) algorithm served as quality metric.}\label{tab:ICP-phantom}
    \begin{tabular}{|c|c|c|}
        \hline
        Scan & \textit{Tattoo} FRE [mm] & Baseline FRE [mm]  \\
        \hline
        1  & 0.67 & 0.88  \\
        2  & 0.57  & 0.84   \\
        3  & 0.66  & 0.89   \\
        \hline
    \end{tabular}
\end{table}

\FloatBarrier
\paragraph{\textit{In vivo} validation:}%+++++++++++++++++
\FloatBarrier

As it was not possible to identify a suitable vessel structure for one of the volunteers, our results are restricted to the 20 scans of the remaining two volunteers, referred to as v1 and v2. As summarized in Table~\ref{tab:in_vivo}, the \textit{Tattoo}-based registration is slightly more reproducible compared to the optical tracking-based method. 

\begin{table}[h!]
\centering
\caption{\textit{In vivo} validation results using a slice-based MVD for both the proposed approach (\textit{Tattoo}) and the optical tracking-based method (baseline).} \label{tab:in_vivo}
    \begin{tabular}{|c|c|c|}
        \hline
        Volunteer   & \textit{Tattoo}  MVD [mm]  & Baseline MVD [mm] \\
        \hline
        v1 & 0.63 & 1.40 \\
        v2 & 0.80 & 1.06 \\
        \hline
    \end{tabular}
\end{table}

\FloatBarrier

\section{Discussion} \label{sect:disc}%+++++++++++++++++++++++++++

% Contribution statement and brief summary
In this paper, we presented \textit{Tattoo tomography}, a novel approach to 3D PA image reconstruction from an acquired sequence of 2D PA image slices. The key advantage of the concept is its simplicity; applying a pattern without the need of involved calibration, training of an algorithm, or any additional complex hardware allows smoothly integrating the proposed approach into any clinical workflow. As the clinically relevant information on the tissue and the information needed to decode the probe pose can be encoded in different wavelengths, the approach does not come at the cost of reduced image quality.  Our prototype implementation of the proposed concept further illustrates that such a pattern-based concept may also feature higher accuracy and precision compared to widely used techniques that rely on external tracking devices. 

% Discussion of results
It is worth mentioning that we had initially planned to use the optical tracking-based approach as reference method. First experiments, however, suggested that the proposed approach might be even more accurate and precise than this method, which is why we decided on a comparative assessment. The FRE obtained for both reconstruction methods may appear relatively high (~0.6-0.8 mm). We attribute this to the fact that the model representation of the phantom was only an approximation of reality. This hypothesis is supported by the fact that both reconstruction methods feature a systematic mismatch of the outer wires, as shown in Fig.~\ref{fig:ICP}.

In our \textit{in vivo} experiment we examined the precision of \textit{Tattoo tomography} in comparison to the optical tracking baseline. The fact that the \textit{Tattoo} approach was substantially more precise then the baseline method in terms of the MVD may most likely be attributed to the different underlying world coordinate systems. The tattoo coordinate system is a relative coordinate system that moves with the target structure and is, therefore, influenced by patient motion to a lesser degree. Due to the limited number of volunteers and the lack of a reliable reference for the reconstruction, further studies are required to support these initial results. 

% Discussion of the method
While the general \textit{Tattoo} concept proposed is potentially very powerful, our first prototype implementation comes with several limitations. The main current drawback is the fact that we use a 2D pattern. A 3D pattern, allowing for the recovery of 6 DoF from a single measurement, would eliminate the need for holding the probe orthogonally to the pattern. The second current limitation, namely the required flatness of the target surface, could then be addressed by image processing-based methods to approximate the shape of the tissue of interest.

Finally, we are planning on enhancing the optical pattern, such that it can not only be used for 3D reconstruction but also for multi-modal registration with conventional imaging data, such as computed tomography (CT) or magnetic resonance imaging (MRI) data. 

% Conclusion
In conclusion, we have presented the first 3D PAI reconstruction concept that elegantly leverages the unique capabilities of PAI. The proposed concept is simple to apply, does not require additional devices such as a tracking system, and can easily be integrated into clinical workflows. According to initial results, it may even be more accurate and precise than competing methods that rely on complex hardware setups. We therefore assume that the concept has high potential for future clinical translation.

\section*{Declarations}

\paragraph{Funding:} This project has been funded by the Data Science Driven Surgical Oncology Program of the National Center for Tumor Diseases (NCT) Heidelberg and received funding from the European Unions Horizon 2020 research and innovation program through the ERC starting grant COMBIOSCOPY under Grant Agreement No. ERC-2015-StG-37960.

\paragraph{Conflicts of interest:} The authors declare that they have no conflict of interest.

\paragraph{Ethics approval:}  The healthy human volunteer experiments were approved by the ethics committee of the medical faculty of Heidelberg University under reference number S-451/2020 and the study is registered with the German Clinical Trials Register under reference number DRKS00023205.

\paragraph{Availability of data/code and material:} Available upon request.

\end{document}